\documentclass{article}
\usepackage{spconf,amsmath,graphicx}
\usepackage{array}
\usepackage{mdwmath}
\usepackage{mdwtab}
\usepackage{eqparbox}
\usepackage{multirow}
\usepackage{stmaryrd}
\usepackage{paralist}
\usepackage{dsfont}
\usepackage{verbatim}
\usepackage{calc}
\usepackage{amssymb}
\usepackage{amstext}
\usepackage{amsthm}
\usepackage{multicol}
\usepackage[small]{caption}
\usepackage{xcolor}
\usepackage{rotating}
\usepackage{subcaption}

\usepackage{algorithm}
\usepackage{algorithmic}
\usepackage[numbers,sort&compress]{natbib}

\usepackage{hyperref}

\usepackage{xcolor}
\usepackage{pgfplots,pgfplotstable}
\usepackage{tikz}
\usepackage{lipsum,adjustbox}
\usetikzlibrary{calc,positioning}
\pgfplotsset{compat=newest}
\newcommand{\ie}[0]{\textit{i.e.},~}
\newcommand{\eg}[0]{\textit{e.g.},~}

\newcommand{\etal}[0]{\textit{et al.}~}

\def\L{{\cal L}}

\title{Towards Generating Ambisonics Using Audio-Visual Cue For \\Virtual Reality}

\name{Aakanksha Rana$^*$\thanks{*These authors contributed equally to this work.}, Cagri Ozcinar$^*$, and Aljosa Smolic}
\address{V-SENSE, School of Computer Science and Statistics, Trinity College Dublin, Ireland.}

\begin{document}
\ninept

\maketitle

\begin{abstract}
Ambisonics \ie a full-sphere surround sound, is quintessential with $360^\circ$ visual content to provide a realistic virtual reality (VR) experience. While $360^\circ$ visual content capture gained a tremendous boost recently, the estimation of corresponding spatial sound is still challenging due to the required sound-field microphones or information about the sound-source locations. In this paper, we introduce a novel problem of generating Ambisonics in $360^\circ$ videos using the audio-visual cue. With this aim, firstly, a novel $360^\circ$ audio-visual video dataset of 265 videos is introduced with annotated sound-source locations. Secondly, a pipeline is designed for an automatic Ambisonic estimation problem. Benefiting from the deep learning based audio-visual feature-embedding and prediction modules, our pipeline estimates the 3D sound-source locations and further use such locations to encode to the B-format. To benchmark our dataset and pipeline, we additionally propose evaluation criteria to investigate the performance using different $360^\circ$ input representations. Our results demonstrate the efficacy of the proposed pipeline and open up a new area of research in $360^\circ$ audio-visual analysis for future investigations.

\end{abstract}
\section{Introduction}
\label{sec:Intro}
Recent advancements in virtual reality (VR) technologies have paved the way of capturing and sharing omnidirectional videos~(ODVs) over social media. ODV, also known as $360^\circ$ video, provides the visual representation of the $360^\circ$ surrounding of the captured scene. This emerging representation can be navigated with three degrees of freedom by rotating and changing the viewing direction of VR devices (\eg tablet, laptop, head-mounted display). Users can nowadays, easily capture the $360^\circ$ content with the help of affordable $360^\circ$ cameras available in the market (\eg Ricoh Theta~\cite{theta360} and Samsung Gear 360~\cite{samsung}) and share their ODVs over social networks to engage viewers deeply. 

Essentially, creating realistic VR experiences requires the ODVs to be captured with their spatial audios. The spatial aspect of sound plays a significant role in informing the viewers about the location of objects in the $360^\circ$ environment, providing an immersive multimedia experience. In practice, however, existing affordable $360^\circ$ cameras can capture the visual scene either with mono or stereo audio signals. As a consequence, such audio-visual content is incapable of creating a magical sense of ``being there" in the VR environment.

Recent user studies have amplified the need for spatial audio to achieve presence in VR films~\cite{Fearghail2018, Knorr2018, Ozcinar2018}. A spatial audio signal (also referred to as 3D audio or 360 audio) is considered as a powerful way of directing viewers' attention~\cite{Dwyer2018, Sitzmann_2018}, however, requiring expensive sound-field microphones, professional sound recording and production tools~\cite{frank2015producing, lim2015}. 

\begin{figure*}[ht]
\centering
		\includegraphics[width=0.9\linewidth]{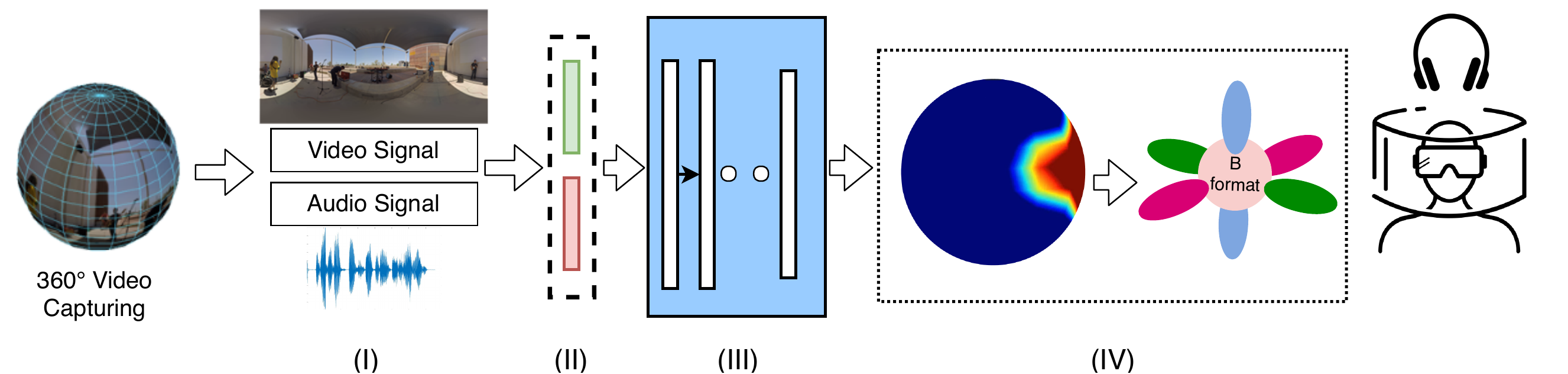}
	\caption{Generation of a full-sphere surround sound environment in ODVs using audio-visual embeddings. The four modules of the pipeline are (I) Representation, (II) Feature embedding, (III) Prediction and (IV) Ambisonics encoding. }\label{fig:pipeline}
\end{figure*}

The lack of spatial audio captured by affordable $360^\circ$ cameras poses an interesting alternative audio-visual research topic in multimedia signal processing community -- \textit{given ODV with a mono/stereo audio signal, can we create the spatial audio to be used for VR video applications?}.

In this work, we introduce a novel problem of generating Ambisonics from the mono/stereo audio signal based on the audio-visual cue. Ambisonics is an effective way of representing the spatial audio and providing 3D sound for VR applications~\cite{enda}. The Ambisonic technology is based on the spherical harmonic decomposition of the sound field and can encode the wave equation in the spherical coordinate system ($r$, $\Phi$, $\theta$). In this context, $r$ is the distance to the source point from the center of coordinates, $\Phi\in(-\pi, \pi]$ and $\theta\in[-\pi/2, \pi/2]$. This direction of the sound can be encoded into four different channels ($W$, $X$, $Y$, $Z$) of the B-format~\cite{Gerzon1985hv}, which is the basis for the first order Ambisonics. Hence, the location of a sound source on the sphere is required to be known to generate Ambisonics.

To tackle this problem, in this paper, we establish a well-annotated dataset and design a 4-stage pipeline to estimate the locations of sound sources. The dataset contains 265 ODVs with different audio-visual scenarios, such as round-table discussions, presentations, meetings, documentaries. In addition, we design a 4-stage pipeline to estimate the locations of sound sources on the sphere to generate the Ambisonics, where the 4-stages are namely, representation, feature embedding, prediction, and encoding. Our proposed pipeline adopts the deep learning based audio-visual feature embeddings and prediction techniques to facilitate the 3D localization of sound source using different ODV input representations. Also, we benchmark our dataset by proposing evaluation criteria to investigate the performance of our pipeline.

In a nutshell, the main contributions of this paper are threefold. First, we address the problem of automatic spatial audio estimation based on audio-visual cue as a first work. Second, we establish the first $360^\circ$ Audio-Visual Dataset (360AVD) which contains 265 video clips with a well-annotated ground-truth providing the sound direction and location. Third, we propose evaluation criteria and perform preliminary quantitative and qualitative analysis using two widely used ODV projection techniques and state-of-the-art feature embedding and prediction algorithms to estimate the location of the sound source. We expect that releasing this dataset and addressing this novel research question will foster further research in multimedia signal processing area.





The rest of this paper is organized as follows. Section~\ref{related} presents the related work on audio-visual machine learning and generating spatial audio. The proposed audio-visual dataset and the proposed pipeline are described in Section~\ref{dataset} and~\ref{sec:Model}. Section~\ref{sec:Results} presents the experiments with the used metrics. Our conclusions are drawn in Section~\ref{conclucion}.

\section{Related Work}

Recent works showed that the location of a sound source could be estimated based on audio-visual signals. For instance, Owens~\etal~\cite{owens_2018_audio} modeled the visual and auditory signals and predicted the sound source pixel location on a given traditional 2D video. Similarly, Tian~\etal~\cite{tian_2018_eccv} proposed audio-guided visual attention mechanism to explore audio-visual correlation with a target to predict event localization in traditional 2D videos. Also, Ephrat~\etal~\cite{Ephrat2018zq} presented a deep network-based model that incorporate audio-visual cue to extract each speaker sound from a mixture of sounds. Audio-visual salient event detection was studied in~\cite{Koutras2015} based on visual, audio and text modalities. The work showed that the performance of visual saliency estimation could be improved by incorporating audio and text. Coleman~\etal~\cite{Coleman2018vs} showed that an object-based audio capturing could achieve a convincing 3D audio experience over headphones. Furthermore, in~\cite{perotin2018}, source separation system was presented for high order Ambisonics recording. A multi-channel spatial filter was derived based on the long short-term memory recurrent neural network with an assumption of known the directions of arrival of the directional sound sources. 

However, to the best of our knowledge, no research on Ambisonics generation based on the audio-visual cue of ODV content currently exists.

\label{related}
\section{Dataset}\label{sec:DS}
\begin{figure}[h]
	\captionsetup[subfigure]{labelformat=empty}
	\begin{subfigure}[b]{.01\textwidth}
		\raisebox{.3in}{\rotatebox[origin=t]{90}{\footnotesize Presentation}}   
	\end{subfigure}
	\begin{subfigure}[b]{.224\textwidth}
		\includegraphics[width=\linewidth]{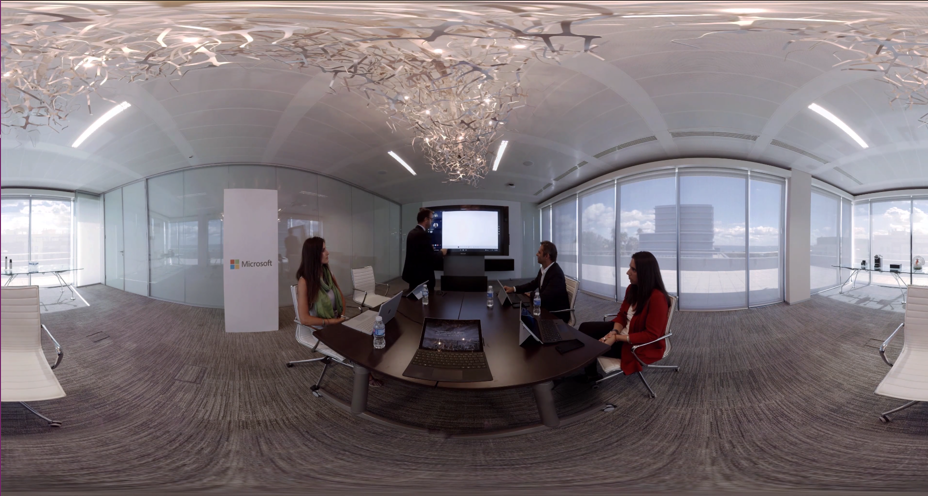}
	\end{subfigure} 
	\begin{subfigure}[b]{.224\textwidth}
		\includegraphics[width=\linewidth]{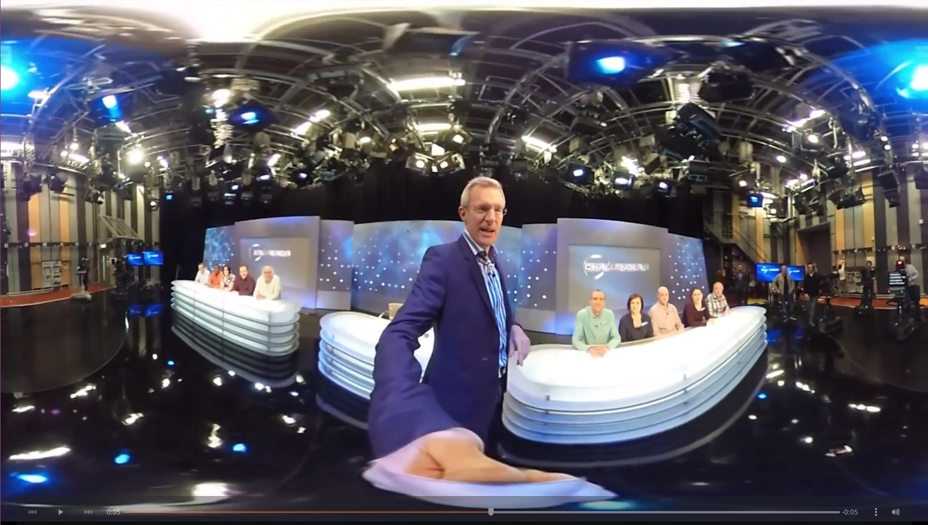}
	\end{subfigure}

	\begin{subfigure}[b]{.01\textwidth}
		\raisebox{.3in}{\rotatebox[origin=t]{90}{\footnotesize Documentary}}   
	\end{subfigure}	
	\begin{subfigure}[b]{.224\textwidth}
		\includegraphics[width=\linewidth]{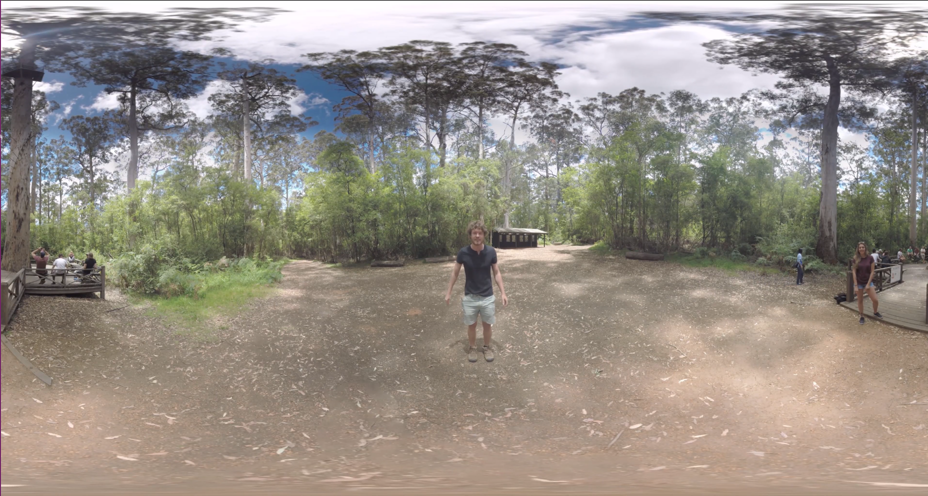}
	\end{subfigure}
	\begin{subfigure}[b]{.224\textwidth}
		\includegraphics[width=\linewidth]{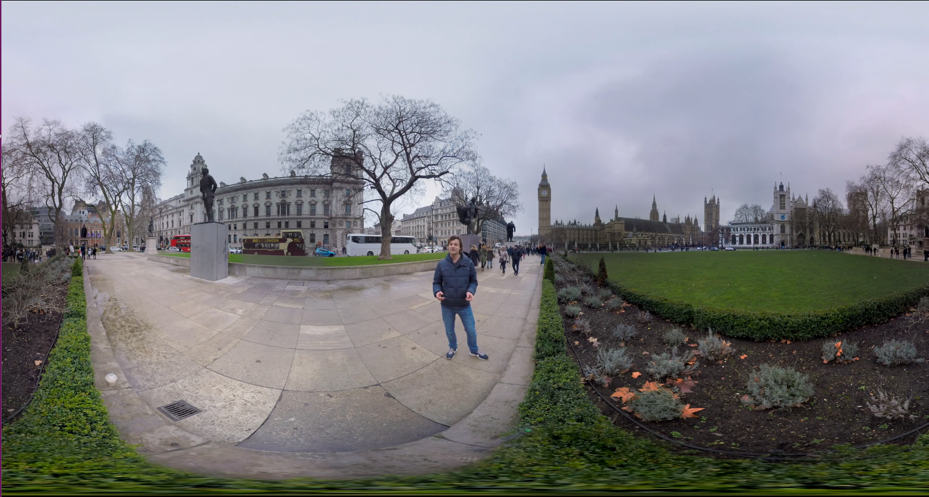}
	\end{subfigure} 
	
	\begin{subfigure}[b]{.01\textwidth}
		\raisebox{.3in}{\rotatebox[origin=t]{90}{\footnotesize Discussions}}   
	\end{subfigure}
	\begin{subfigure}[b]{.224\textwidth}
		\includegraphics[width=\linewidth]{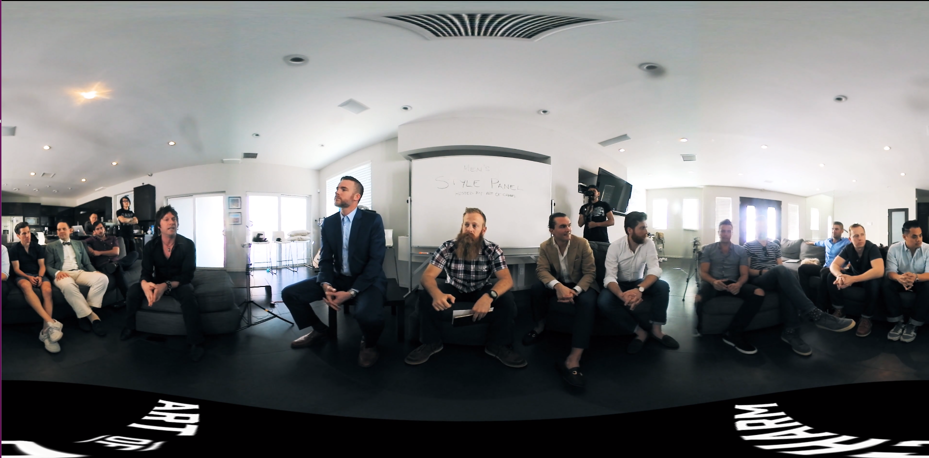}
	\end{subfigure} 
		\begin{subfigure}[b]{.224\textwidth}
		\includegraphics[width=\linewidth]{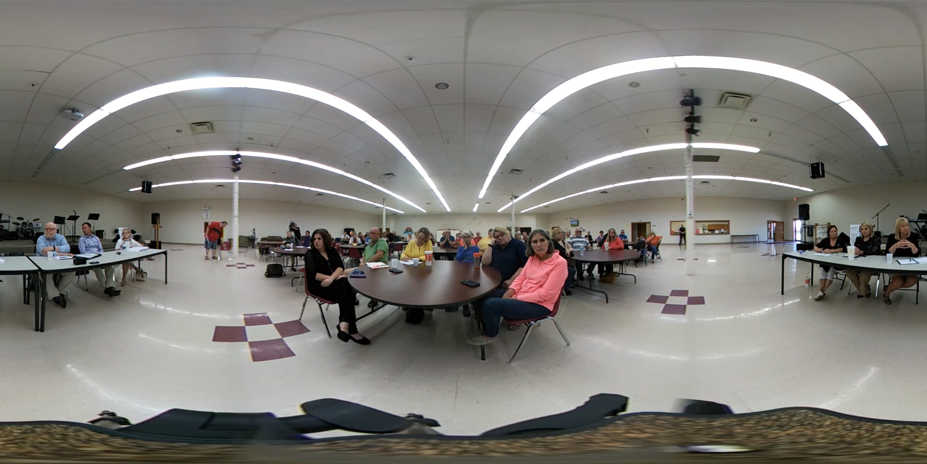}
	\end{subfigure} 
	
	\begin{subfigure}[b]{.01\textwidth}
		\raisebox{.3in}{\rotatebox[origin=t]{90}{\footnotesize Debates}}   
	\end{subfigure}
	\begin{subfigure}[b]{.224\textwidth}
		\includegraphics[width=\linewidth]{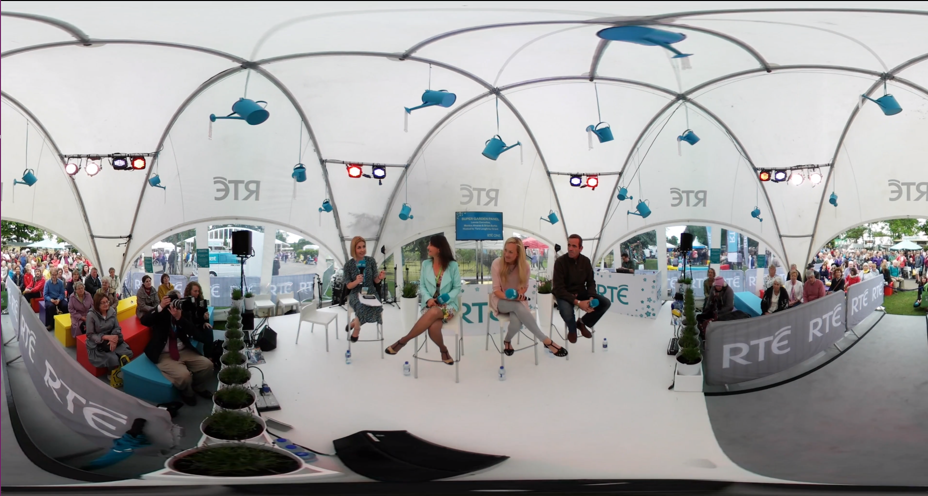}
	\end{subfigure} 
	\begin{subfigure}[b]{.224\textwidth}
		\includegraphics[width=\linewidth]{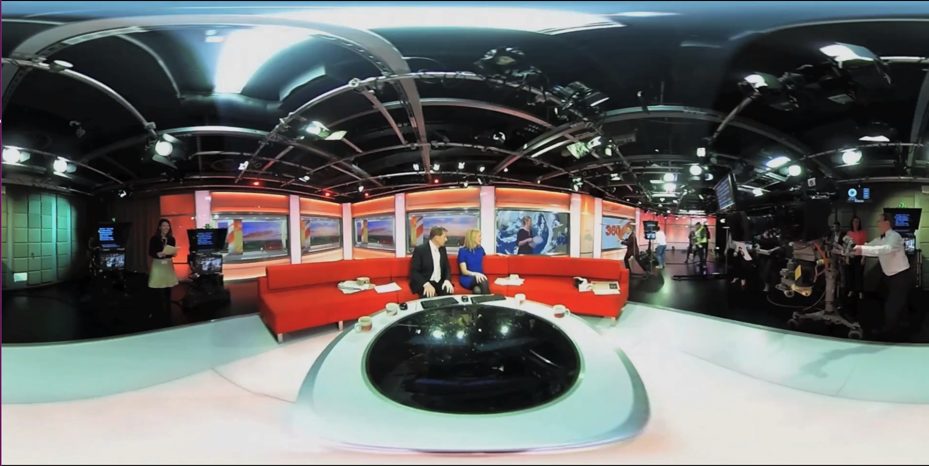}
	\end{subfigure}
	\caption{Sample scenes from the proposed 360AVD.}\label{fig:dataset}
\end{figure}

Dataset is fundamental in predicting the sound location on ODV. However, to the best of our knowledge, there is no publicly available dataset suitable for our objective. To this end, we present the first 360AVD which contains 265 video clips with a well-annotated ground-truth providing the sound direction and location. The dataset has been prepared using publicly available YouTube $360^{\circ}$ unlabeled videos. Each clip in the dataset is of $10$ seconds where all the audio sources are manually annotated per second. To annotate the audio source locations, we used Microsoft's visual object tagging tool\footnote{Visual Object Tagging Tool: An electron app for building end to end Object Detection Models from Images and Videos: \url{https://github.com/Microsoft/VoTT}} which supports labeling of multiple pixel locations for each second of a given video content. The dataset contains a different range of categories: presentation, documentary, debates and casual discussions. Sample scenes from the proposed dataset are presented in Figure~\ref{fig:dataset}. Our proposed dataset with our source codes are available at~\url{https://github.com/V-Sense/360AudioVisual}.


\label{dataset}
\section{Proposed system}\label{sec:Model}
Figure~\ref{fig:pipeline} illustrates the proposed pipeline for the generation of a full-sphere surround sound environment in ODV. The proposed system consists of four main modules, namely (I) input representations, (II) feature embedding, (III) prediction models, and (IV) Ambisonics encoding. At the first module, our aim is to investigate the impact of different sphere-to-planar projections for ODV. At the second module, we jointly estimate features for audio and video signals. Then, the combined information from the video and audio signals is fed into the prediction module to predict the 3D sound source location of a given ODV. Finally, we generate first-order Ambisonics by including the direction of the sound and encoding the estimated multi-channel audio based on the B-format.

We detail each module of the pipeline in the following sub-sections.

\subsection{Input Representations}
\begin{figure}[h]
\centering
	\captionsetup[subfigure]{labelformat=empty}
	\begin{subfigure}[b]{\linewidth}
		\includegraphics[width=\linewidth]{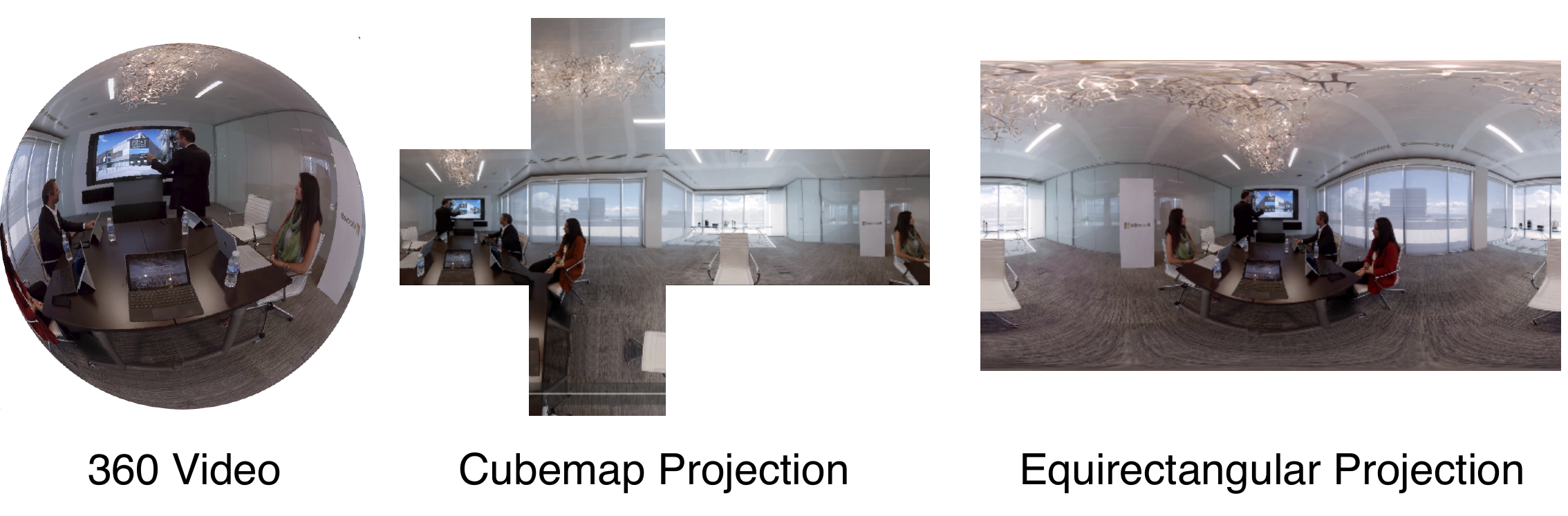}
	\end{subfigure} 
	\caption{Visual representation of equirectangular and cubemap projections for the captured $360^\circ$ video.}\label{fig:formats}
\end{figure}

We investigate the performance of two widely used ODV representations, equirectangular and cubemap (or cubical), as shown in Figure~\ref{fig:formats}. The first, an equirectangular projection, is the most straightforward format that represents a spherical object on a 2D planar surface. The second, a cubic projection, is a collection of six cube faces which are utilized to fill the whole sphere around the viewer. The first projection, however, contains less geometrical distortions than the second one.

\subsection{Feature Embedding}\label{prediction}
Combining visual~\cite{rana14} and audio embedding using deep learning techniques have been recently studied in the literature~\cite{owens_2018_audio,tian_2018_eccv} for several end-to-end application-based scenarios such as source separation, action recognition, and audio-visual alignment. Most of the works are designed for traditional 2D video with a mono/stereo audio signal. However, such models combining the visual and sound information have not been studied for $360^\circ$ video content.

Motivated from~\cite{owens_2018_audio,tian_2018_eccv}, we formulate the feature embedding and prediction modules of the proposed Ambisonics generation pipeline. For the feature embedding module, we firstly employ the pre-trained VGG-19~\cite{SimonyanZ14a} network to compute the visual features using the selected $360^\circ$ visual representation \ie equirectangular or cubical projections. For each second, similar to~\cite{tian_2018_eccv}, we compute the feature maps from $15$ frames and average them to obtain one feature map, $v$, of the dimension of $7\times7\times512$. For each second of audio signal, we simultaneously extract the $128$ dimensional audio representation, $a$, using a pre-trained VGG-like network~\cite{HersheyCEGJMPPS16}. Then, we finally feed the extracted feature embedding to the following prediction module to obtain the sound source probability maps, as shown in Figure~\ref{fig:pipeline}. 

\begin{table*}[ht]
	\centering{
		\begin{tabular}{c|ccc|ccc}
			\hline
			\multirow{2}{*}{\textbf{Models}} & \multicolumn{3}{c}{\textbf{360-SSD}} & \multicolumn{3}{c}{\textbf{360-OvErr}} \\	
 			\cline{2-7}
             & $\epsilon$=0.6 & 0.5  &  0.4 & 0.6 & 0.5 & 0.4  \\ 
			\hline
			SsM-Cubical	& \textbf{0.71 $\pm$ 0.04} & 0.72 $\pm$ 0.08  & 0.74 $\pm$ 0.06 &  \textbf{0.71 $\pm$ 0.06}
			&  0.77 $\pm$ 0.05 & 0.82 $\pm$ 0.04 \\
			SsM-EquiR & 0.75 $\pm$ 0.06 & 0.77 $\pm$ 0.09 & 0.79  $\pm$ 0.07 & 0.78 $\pm$ 0.07 & 0.84 $\pm$ 0.06 & 0.88 $\pm$ 0.08 \\

			Att-Cubical	& 0.72 $\pm$ 0.05 & 0.73 $\pm$ 0.05  & 0.74 $\pm$ 0.04 & 0.72 $\pm$ 0.05 & 0.74 $\pm$ 0.08 & 0.78 $\pm$ 0.08  \\
			Att-EquiR & 0.76 $\pm$ 0.04 & 0.77 $\pm$ 0.08  & 0.78 $\pm$ 0.06 & 0.84 $\pm$ 0.06 & 0.85 $\pm$ 0.06 & 0.86 $\pm$ 0.06  \\			
			\hline
		\end{tabular}
	}    
	\caption{\footnotesize Quantitative Results on 360AVD Dataset. The scores are averaged on 265 ODVs for all models.}
	\vspace{-5 mm}
	\label{tab:quantresults}
\end{table*}

\subsection{Prediction Module}
To predict the location of the sound source, we first adopt the sub-networks of the proposed deep networks, Tian~\etal~~\cite{tian_2018_eccv} and Owens~\etal~\cite{owens_2018_audio}, originally designed for traditional 2D videos. We then alter them for our task at hand \ie to predict the 3D volumetric maps. Both models are recently published and their models are publicly available. We used the middle-layers from these pre-trained models to obtain the sound source location maps. The two prediction modules of our pipeline are named as a self-supervised module (SsM) and attention module (Att). 

\textbf{SsM module}: It is an adaptation of the fusion sub-network proposed in~\cite{owens_2018_audio} with three convolutional layers. To predict the sound localization map $S_p^{SsM}$, we concatenate the feature embedding and feed it to the convolutional layers, and finally apply the spherical mapping function $f$ over the probability estimation map given as:

\begin{equation}
    S_p^{SsM} = f(\sigma(\L^Tconv_l)),
\end{equation}
where $\sigma$ is the sigmoid function, $\L$ is the affine layer, $conv_l$ is the last convolutional layer and function $f$ maps $ (x,y) \rightarrow (\theta,\phi)$ coordinates. 

\textbf{Att module}: It is inspired from attention mechanism detailed in~\cite{tian_2018_eccv}, adaptively learns to locate the visible regions in each second of the video from where the sound originates. To predict 3D sound for each second, localization maps $S_p^{Att}$ is mathematically defined as:

\begin{equation}\label{eq:att}
    S_p^{Att} = f(softmax(\omega \cdot \rho(l_v) + l_a)),
\end{equation}
where the $\rho_r$ is a hyperbolic tangent function, $\omega$ is a weighting parameter and $l_a$ and $l_v$ are the audio-visual transformation layers. 
Altogether, the attention weight vector computed using the multi-layer perception (MLP) like formulation as detailed in~\cite{tian_2018_eccv}, is finally transformed by applying function $f$.

\subsection{Ambisonics Encoding}
After the sound localization maps $S$ is estimated, we encode localized sound sources to the B-format. For this, the location of the $i$-th sound source is first estimated as follows:
\begin{equation}
\tilde{\Phi_i}, \tilde{\theta_i} =  \mathcal{C}    
\end{equation}
where $\mathcal{C}$ is the set of a sound source location based on 3D coordinates, $\{{C}\}_{i=1}^{N}$, $N$ is the number of sound sources, $\tilde{\Phi_i}$ and $\tilde{\theta_i}$ are the predicted spherical locations of the $i$-th sound source. The center of sound source $C_i$ is the mean location of distribution of 3D point in the $i$-th sound source probability volumes $S$. The spherical volumes $S$ are obtained by an absolute threshold, \ie for all coordinates where $S(x,y,z) \leq \epsilon$, values are equated to $0$. Hence, we encode the B-format as follows:

\begin{equation}
\begin{aligned}
\begin{split}
\centering
    W(t) &= \sum_i^N s_i(t)/\sqrt{2},\\
    X(t) &= \sum_i^N s_i(t)cos \tilde{\Phi_i} cos \tilde{\theta_i},\\
    Y(t) &= \sum_i^N s_i(t)sin \tilde{\Phi_i} cos \tilde{\theta_i},\\
    Z(t) &= \sum_i^N s_i(t)sin \tilde{\theta_i},
\end{split}
\end{aligned}
\end{equation}
where $s_i(t)$ is the $i$-th sound signal of a given ODV, and the set of four audio channels $(W,X,Y,Z)$ form the estimated Ambisonics. The non-directional sound pressure level is represented as $W$, and three other channels, $(X, Y, Z)$, are described as the position of the sound: front-to-back ($X$), side-to-side ($Y$), and up-to-down ($Z$). 

\label{method}
\section{Experiments}\label{sec:Results}
This section describes the proposed metrics and preliminary results obtained from our proposed pipeline for Ambisonics generation.
\subsection{Metrics}
To evaluate the performance of the predicted sound source location quantitatively, we introduce two metrics, namely, 360 Sound Source Distance (360-SSD) and 360 overlap error (360-OvErr).

\textbf{360-SSD:} It estimates the Euclidean distance between the centre of the predicted $i$-th sound source, $C^p_i (x,y,z)$, and the centre of ground truth $i$-th sound source, $C^g_i (x,y,z)$, in the Cartesian coordinate system \ie $||C^g_i - C^p_i||$. The center of sound source $C_i$ is defined in Section 4.3. For 360-SSD, all distances are normalized and the probability spheres have radius 0.5.

\textbf{360-OvErr:} This metric is based on the ratio of an intersection of the predicted and ground truth probability volumes to the union and, mathematically given as $ 1- \frac{S_p \cap S_g}{S_g \cup S_p}  $, where $S_p$ is the predicted probability volumes and $S_g$ belongs to the ground truth. This measure can be seen as a 3D variant of single object localization error proposed in~\cite{Russakovsky_2015_IJCV}.

\subsection{Implementation}
\textbf{Feature embedding and prediction.} We base on the VGG-19~\cite{SimonyanZ14a} and VGG-like~\cite{HersheyCEGJMPPS16} network for feature embedding module and computing audio and visual features are trained on Imagnet~\cite{SimonyanZ14a} and Audioset~\cite{Gemmeke} dataset. For our SsM module, we adopt the convolutional layers from~\cite{owens_2018_audio} which are trained large scale dataset of 750,000 videos. Similarly, we adopt the transformation layers $l_a$ and $l_v$ in Eq.~\ref{eq:att} from the model proposed in~\cite{tian_2018_eccv} trained on a large-scale AVE dataset. The followed training paradigm is similar to~\cite{owens_2018_audio,tian_2018_eccv}.\\
\textbf{Ambisonic encoding.} To encode Ambisonics, we used the Facebook 360 encoder tool from the Facebook Spatial Workstation~\cite{fb360}. Each predicted location for each sound was added to the location channels of B-format. Afterward, the MP4Box~\cite{mp4box} was used to wrap the ODV and multi-channel audio together within an MP4 header file.

\subsection{Results}
In this section, we carried out the performance evaluation study of the proposed Ambisonics generation pipeline using both well-known representations and state-of-the-art prediction models over the proposed 360AVD dataset. 

We first evaluated the performance of the predicted probability volumes, quantitatively, by using the proposed metrics: 360-SSD and 360-OvErr. For both metrics, less score stands for better performance. In Table~\ref{tab:quantresults}, we present the results averaged over 265 ODVs with cubical and equirectangular formats and different settings of $\epsilon$. Using both the metrics, we observe that SsM model-based prediction module with cubical ODV input representation performs the best in terms of localizing the exact sound source 3D location as well as the region. Att model-based prediction module competes closely with the former for localizing the 3D sound source. 

\begin{figure}[h]
	\captionsetup[subfigure]{}
	\begin{subfigure}[b]{0.157\textwidth}
		\includegraphics[width=\linewidth]{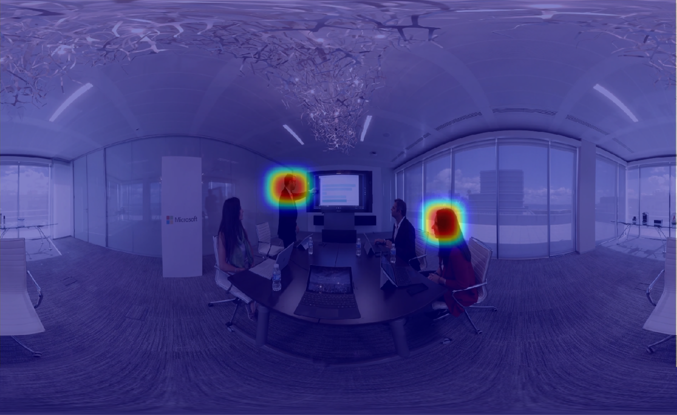}
		\caption{Original EquiR}
	\end{subfigure}
	\begin{subfigure}[b]{0.157\textwidth}
		\includegraphics[width=\linewidth]{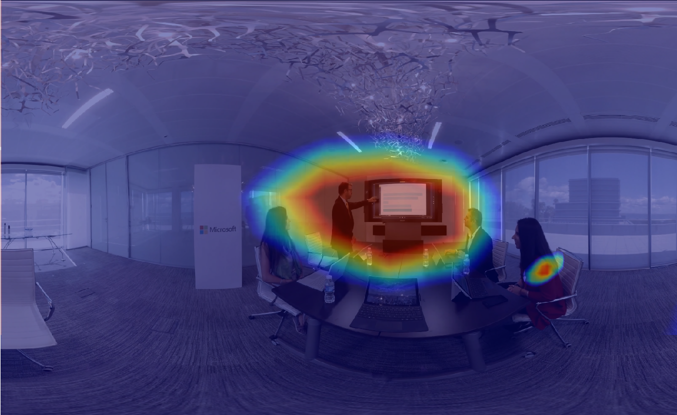}
		\caption{SsM-EquiR}
	\end{subfigure}
	\begin{subfigure}[b]{0.157\textwidth}
		\includegraphics[width=\linewidth]{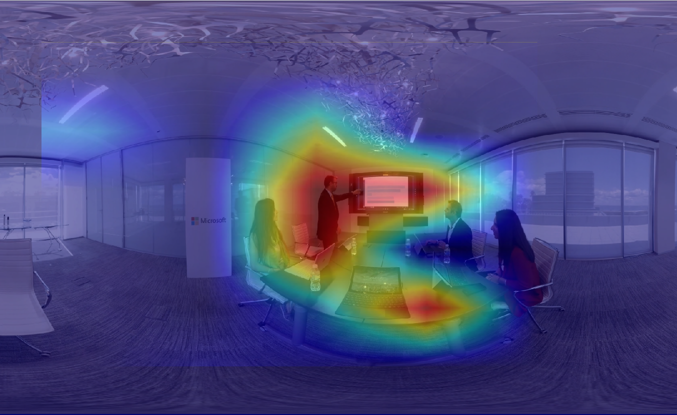}
		\caption{Att-EquiR}
	\end{subfigure}
	
	\begin{subfigure}[b]{0.157\textwidth}
		\includegraphics[width=\linewidth]{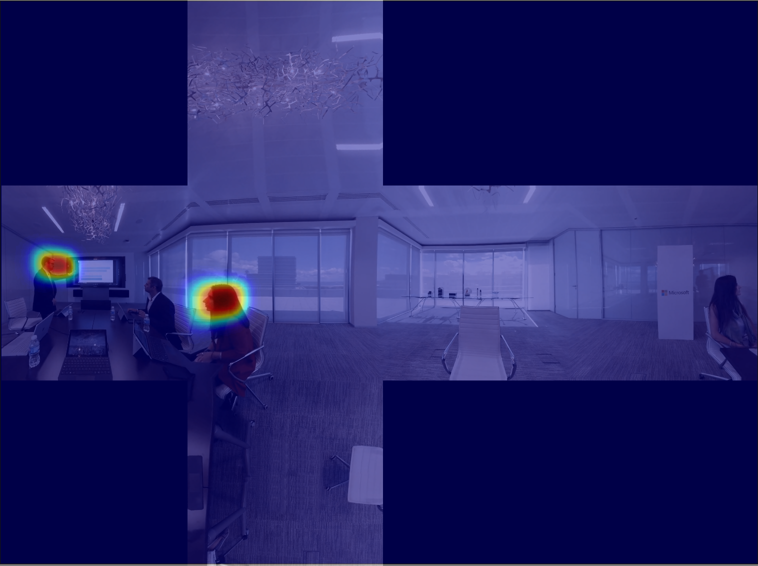}
		\caption{Original-Cubical}
	\end{subfigure}
	\begin{subfigure}[b]{0.157\textwidth}
		\includegraphics[width=\linewidth]{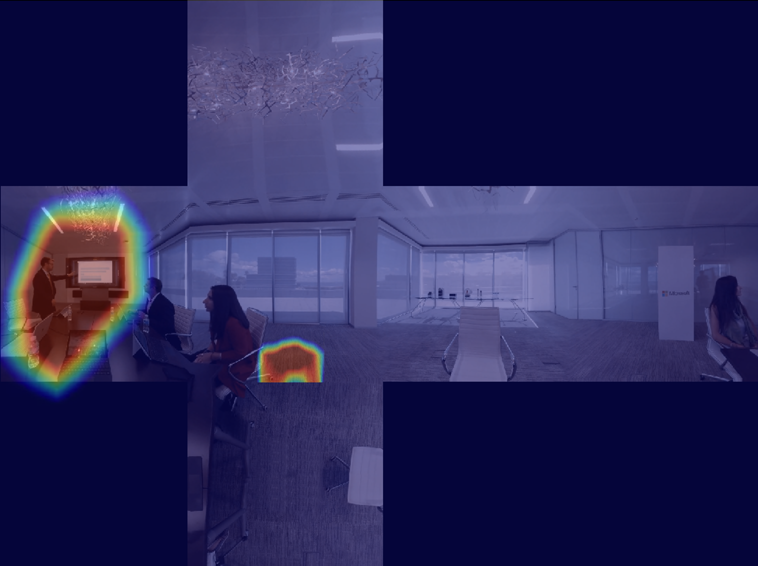}
		\caption{SsM-Cubical}
	\end{subfigure}
	\begin{subfigure}[b]{0.157\textwidth}
		\includegraphics[width=\linewidth]{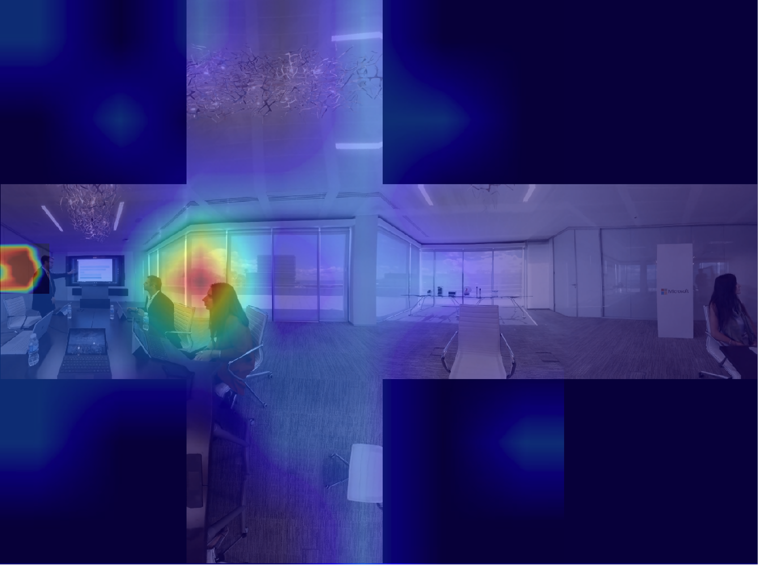}
		\caption{Att-Cubical}
	\end{subfigure}
	\caption{Qualitative Results: Row I shows the original $360^\circ$ video frame with (a) overlay-ed ground truth, predicted results from (b) SsM and (c) Att modules in equirectangular representation. Row II shows the same frame with (d) overlay-ed ground truth, predicted results from (e) SsM and (f) Att modules in cubical representation.}\label{fig:Qresults}
\end{figure}

For both models, we observe that cubical ODV representation provides better localization than the equirectangular format. This partly accounts to less distortions present in the cubical format, which in turn is favorable for distinctive feature embedding and prediction. On the other hand, the overall higher 360-OvErr with equirectangular format demonstrate the influence of the distortions where it leads to larger detected regions. This can be additionally seen in Fig.~\ref{fig:Qresults} where the qualitative performance of both models are illustrated with cubical and equirectangular representation. Additional results, \eg ODV with generated Ambisonics, are available on our GitHub page.

\label{result}
\section{Conclusion}\label{sec:Conc}

This paper introduces a novel research problem of automatic Ambisonics generation from the mono/stereo audio signal based on audio-visual cue. For this aim, we propose a pipeline to predict the sound source location in a 3D space and time. The proposed pipeline contains four stages, representation, feature embedding, prediction, and Ambisonics encoding. To investigate the performance of each module, we introduce the first audio-visual dataset of 265 omnidirectional videos consisting of various single to multiple speech scenarios and evaluation metrics. Our initial analysis shows that the cubical representation of omnidirectional video with the self-supervised deep learning prediction algorithm performs better performance. All obtained results suggest that the problem of accurate sound source location estimation using audio-visual cue for Ambisonics remains open with a quite large room of improvement. The future work will consider developing optimal modules for our end-to-end pipeline for Ambisonics generations.


\label{conclucion}
\footnotesize
\section*{Acknowledgements}
\footnotesize{This publication has emanated from research conducted with the financial support of Science Foundation Ireland (SFI) under the Grant Number 15/RP/2776. We gratefully acknowledge the support of NVIDIA Corporation with the donated GPU used for this research. We gratefully acknowledge Dr. Enda Bates for all the insightful discussions and help.}

\footnotesize
\bibliographystyle{IEEEbib}
\bibliography{references}

\end{document}